\documentclass[aps,preprint,showpacs,superscriptaddress]{revtex4}
\usepackage{amsfonts}
\usepackage{amsmath}
\usepackage{amssymb}
\usepackage{graphicx}%
\begin{document}
\title{Thermodynamics of acoustic black holes in two dimensions}
\author{Baocheng Zhang}
\email{zhangbc.zhang@yahoo.com}
\affiliation{School of Mathematics and Physics, China University of Geosciences, Wuhan
430074, China}

\begin{abstract}
It is well-known that the thermal Hawking-like radiation can be emitted from
the acoustic horizon, but the thermodynamic-like understanding for acoustic
black holes was rarely made. In this paper, we will show that the kinematic
connection can lead to the dynamic connection at the horizon between the fluid
and gravitational models in two dimension, which implies that there exists the
thermodynamic-like description for acoustic black holes. Then, we discuss the
first law of thermodynamics for the acoustic black hole via an intriguing
connection between the gravitational-like dynamics of the acoustic horizon and
thermodynamics. We obtain a universal
form for the entropy of acoustic black holes, which has an interpretation
similar to the entropic gravity. We also discuss the specific heat, and find
that the derivative of the velocity of background fluid can be regarded as a
novel acoustic analogue of the two-dimensional dilaton potential, which
interprets why the two-dimensional fluid dynamics can be connected to the
gravitational dynamics but difficult for four-dimensional case. In
particular, when a constraint is added for the fluid, the analogue of a
Schwarzschild black hole can be realized.

\end{abstract}
\maketitle

\section{Introduction}

The concept of acoustic black holes, based on the kinematical analogue between
the motion of sound wave in a convergent fluid flow and the motion of a scalar
field in the background of Schwarzschild spacetime, had been suggested
initially in 1981 \cite{wgu81}, and then many physical systems had been used
to present the similar concept, such as Bose-Einstein condensates (BEC)
\cite{gaz00}, superfluid He \cite{jv98}, slow light \cite{us03},
electromagnetic waveguide \cite{su05}, light in a nonlinear liquid
\cite{efb12}, i.e. see the review \cite{blv11}. Recently, the observation of
the acoustic horizon had been made in different physical systems
\cite{pkl08,lis10,bcf10,hrc10}. But the Hawking-like radiation \cite{swh74} is
still difficult to be observed due to the small radiation temperature, while
the recent report \cite{js14} on the observation of Hawking-like radiation was
related to a self-amplifying mechanism called as black-hole laser suggested by
Corley and Jacobson \cite{cj99}. Thus the analogue of black holes provides a
prospective avenue to observe the Hawking radiation experimentally, since the
direct observation from astrophysical black holes is nearly impossible due to
the temperature of several orders of magnitude lower than the cosmic microwave background.

As stated above, the acoustic black hole is based on only the kinematical
analogue, irrelevant to the dynamics. But for Schwarszchild black holes, once
the radiation starts, it will get hotter and hotter by losing energy, which is
evident from the relation that the temperature is inversely proportional to
the mass. This description is beyond kinematics, and is mainly based on
thermodynamics that is definitely dependent on the dynamics \cite{mv97,mv98}.
Recently, it was found that when the backreaction \cite{bffp05,bff05} was
considered (some different views for this was seen in Ref.
\cite{ms00,suxf05,ms07}, but how to discuss the backreaction exactly is still
open now \cite{blv11}), the acoustic black hole will not present the similar
thermodynamic behaviors of Schwarzschild black holes while looks more like a
near-extremal Reissner-Nordstr\"{o}m black hole. This is an important step for
the possible thermodynamics of an acoustic black hole. Next, as for
thermodynamics of a gravitational black hole, the description has usually to
be made with the help of proper expressions for the mass and the entropy of
the acoustic black hole. Some recent developments had showed that the entropy
could be endowed to an acoustic black hole by understanding the microscopical
modes in some situations \cite{sg11,mr11,mr13,abs14}. But the mass is
definitely relevant to the classical dynamics, and thus its definition
requires an analogue for gravity's equation. An interesting work \cite{mc05}
had been attempted for this, in which the mass and the entropy were defined by
using an analogue between two-dimensional (2D) dilaton black holes and the
acoustic black holes but at the same time the fluid dynamics had been fixed by
the two-dimensional dilaton gravity \cite{gkv02}. Thus by the analogue for
gravitational dynamics, Cadoni \cite{mc05} realized the thermodynamics for an
acoustic black hole in two dimension.

In the earlier discussion about thermodynamics of acoustic black holes, the
dynamics is fixed in advance. In this paper, we want to study whether the
dynamics of fluid can support the description of thermodynamics spontaneously
only if the acoustic horizon has formed. This reminds us of a method which is
well-known in gravity, that is, Einstein's equation can be derived from the
thermodynamics plus the knowledge from the black hole physics
\cite{tg95,egj06}. Thus, the Einstein's equation could be regarded as a
thermodynamic identity, i.e. see the review paper \cite{tp05}. That means that
at the horizon, the Einstein's equation is equivalent to the thermodynamic
first law, and the thermodynamic quantities such as the mass and the entropy
can be read off directly from the Einstein's equation. Then, one might ask
whether this can be extended to the fluid model, or whether it is permitted to
read off the mass and the entropy of an acoustic black hole from the fluid
equation. It is feasible if the kinematical analogue can give the connection
between two kinds of dynamics. Moreover, the application of the method should
be able to give the consistent results with those obtained by other methods
\cite{bffp05,bff05,mc05}. In this paper, we will try to explore these from the
equation of motion for the fluid, but our discussion is limited in two dimensions. Once if the
mass and the entropy of the acoustic black hole are given, we can estimate the
thermodynamic evolution using the specific heat, as used usually for the
Schwarzschild black hole. By the estimation, we are able to find that how the
elements in a fluid model influence the evolution of the acoustic black hole.

The structure of the paper is as follows. We will first revisit the concept of
the acoustic black hole with the model used in the initial paper of Unruh, and
explain the Hawking-like radiation with the perturbed action. In the third
section we will analyze the background fluid equations, and construct the
kinematical connection between acoustic black holes and 2D black holes, which
leads to the dynamic connection between them. Then we identify the
corresponding thermodynamic quantities by the similar method for the
Einstein's equation. We also discuss the same identification from the
equations for BEC as an acoustic analogue of black holes. The fourth section
contributes to the thermodynamic stability of the acoustic black hole with the
aid of the specific heat. Finally, we discuss and summarize our results in the
fifth section.

\section{Acoustic black hole}

Consider an irrotational, barotropic fluid that was also considered in the
seminal paper \cite{wgu81} of Unruh with the action \cite{bff05},
\begin{equation}
S=-\int d^{4}x\left[  \rho\dot{\psi}+\frac{1}{2}\rho\left(  \overrightarrow
{\nabla}\psi\right)  ^{2}+u\left(  \rho\right)  \right]  \label{abha}%
\end{equation}
where $\rho$ is the mass density, $\psi$ the velocity potential, i.e.
$\overrightarrow{v}=\overrightarrow{\nabla}\psi$, and $u\left(  \rho\right)  $
the internal energy density. In this paper we only involve the linear
perturbation for the derivation of the acoustic metric, so the action
(\ref{abha}) is enough for our discussion.

The variation of $S$ will give the equations of motion, and one of them is the
Bernoulli equation,
\begin{equation}
\dot{\psi}+\frac{1}{2}\overrightarrow{v}^{2}+\mu\left(  \rho\right)  =0,
\label{be}%
\end{equation}
where $\mu\left(  \rho\right)  =\frac{du}{d\rho}$, and the other one is the
continuity equation,%
\begin{equation}
\dot{\rho}+\overrightarrow{\nabla}\cdot\left(  \rho\overrightarrow{v}\right)
=0. \label{ce}%
\end{equation}
We linearize the density and the velocity potential by replacing $\psi
\rightarrow\psi+\phi,\rho\rightarrow\rho+\sigma$ where $\psi$ and $\rho$ are
related to the background fluid determined by the equations of motion, and
$\phi$ and $\sigma$ are small perturbations. When we put the new velocity
potential and mass density after perturbations into the initial action
(\ref{abha}), a new added term, up to quadratic order, appears as,%
\begin{equation}
S_{p}=-\frac{1}{2}\int d^{4}x\left[  \rho\left(  \overrightarrow{\nabla}%
\phi\right)  ^{2}-\frac{\rho}{c_{s}^{2}}\left(  \dot{\phi}+\overrightarrow
{v}\cdot\overrightarrow{\nabla}\phi\right)  ^{2}\right]  , \label{abhp}%
\end{equation}
where the speed of sound $c_{s}$ is defined as $c_{s}^{2}=\rho\frac{d\mu
}{d\rho}$ and the equation of motion for $\phi$, $\phi+$ $\frac{\rho}%
{c_{s}^{2}}\left(  \dot{\phi}+\overrightarrow{v}\cdot\overrightarrow{\nabla
}\phi\right)  =0$, is used. It is noted that for some cases such as presented
in Unruh's original model \cite{wgu81}, a completely appropriate approximation
can be taken for the speed of sound as a position-independent constant, but
there is also some other cases such as BEC \cite{gaz00,blv01} which didn't
take the speed of sound as a constant. In this paper, we take the speed of
sound as constant, in addition to the special case where the non-constant
speed of sound will be pointed out.

The propagation of sound wave on the background fields can be obtained by a
wave equation, $\nabla^{2}\phi=\frac{1}{\sqrt{-g}}\partial_{\mu}\left(
\sqrt{-g}g^{\mu\nu}\partial_{\nu}\phi\right)  $, where $g_{\mu\nu}$ can be
read off from the so-called acoustic metric,%
\begin{equation}
ds^{2}=\frac{\rho}{c_{s}}\left[  -c_{s}^{2}dt^{2}+\left(  d\overrightarrow
{x}-\overrightarrow{v}dt\right)  ^{2}\right]  . \label{abhm}%
\end{equation}
This metric, if considered as a metric of acoustic black hole, can be
understood from a model called as the river model of black holes
\cite{hl04,chs08}: the fluid as the background is flowing along a direction to
the region beyond the Newtonian escape velocity (that is the local velocity of
sound), and the point where the velocity of background fluid equals to the
sound velocity represents the horizon of the black hole. From this
description, it is noted that the essence of the acoustic black hole, in
particular, Hawking radiation which propagates against the fluid, can be
understood with a two-dimensional model. So in the paper we will consider only
the case of two dimensions, for which we take the direction of fluid flowing
as the $x$-axis and thus the components of the speed $\overrightarrow{v}$
along other two directions will be suppressed naturally. Moreover, one can
also reduce the dimensions by the spherical symmetry, in which this analogue
will reproduce the results for the so-called \textquotedblleft dirty black
holes\textquotedblright\ \cite{mv92}.

Furthermore, if the scalar field $\phi$ is quantized, a similar effect to
Hawking radiation appears at the horizon with the temperature which can be
expressed as%
\begin{equation}
T_{a}=\frac{\hbar}{4\pi c_{s}k_{B}}\frac{d(c_{s}^{2}-v^{2})}{dx}|_{v=-c_{s}%
}=\frac{\hbar}{2\pi k_{B}}v^{^{\prime}}\left(  x_{h}\right)  , \label{at}%
\end{equation}
where the horizon is located at $x=x_{h}$ that is determined by $v\left(
x_{h}\right)  =-c_{s}$ and the prime is the derivative with regard to the
coordinate $x$. A better interpretation for the origin of Hawking radiation
can be from the analysis for the corrected action $S_{p}$ \cite{ms07}, in
which some extra terms has to be added in order to ensure the positivity of
perturbation modes and differentiability of the classical ground state profile
(i.e. $\partial_{x}\phi$), and thus the superluminal modes that can overcome
the frame-dragging speed and approach the horizon from the inside will be
ripped apart at the horizon so that the part with positive frequencies escapes
into the exterior region and the others drop again into the interior domain.
This interpretation is consistent with that from the quantization, but it
evidently draws support from the dispersion relation that is necessary to be
considered in the realistic experiment of simulation about black/white hole
horizon and that can also avoid the trans-Planckian puzzle \cite{ms07}.

\section{Thermodynamic description}

At first, let us review how the Einstein equation is identified with the
thermodynamic relation \cite{tp05,tp02,ksp07}. Start with the stationary
metric in ADM form \cite{kw95},%
\begin{equation}
ds^{2}=-N_{t}(r)^{2}dt^{2}+[dr+N_{r}(r)dt]^{2}+r^{2}d\Omega^{2}, \label{me}%
\end{equation}
where $N_{t}(r)$ and $N_{r}(r)$ are the lapse and shift functions,
respectively.\ The metric is well behaved on the horizon, and for a four
dimensional spherically Schwarzschild solution, $N_{t}=1,N_{r}=\sqrt{\frac
{2M}{r}}$ ($M$ is the mass of the black hole); for a four dimensional
Reissner-Nordstr\"{o}m solution, $N_{t}=1,N_{r}=\sqrt{\frac{2M}{r}-\frac
{Q^{2}}{r^{2}}}$ ($M$ is the mass and $Q$ is the charge of the black hole). In
particular, it is noted that the acoustic metric can be obtained by taking
$N_{t}=-c_{s},N_{r}=-v$ up to a conformal factor $\frac{\rho}{c_{s}}$.

For the metric (\ref{me}), the horizon, $r=r_{h}$, is determined from the
condition $N_{t}(r_{h})-N_{r}(r_{h})=0$. The temperature associated with this
horizon is $k_{B}T_{H}=\frac{\hbar c}{4\pi}\frac{d}{dr}\left(  N_{t}^{^{2}%
}(r)-N_{r}^{^{2}}(r)\right)  |_{r=r_{h}}=$ $\frac{\hbar cN_{t}(r_{h})}{2\pi
}\left(  N_{t}^{^{\prime}}(r_{h})-N_{r}^{^{\prime}}(r_{h})\right)  $ where the
prime is the derivative with regard to the coordinate $r$. Consider the ($r$,
$r$) components of the vacuum Einstein equation, $R_{rr}=0$, and evaluate it
at the horizon; this gives%
\begin{equation}
N_{t}(r_{h})\left[  N_{t}^{^{\prime}}(r_{h})-N_{r}^{^{\prime}}(r_{h})\right]
-\frac{1}{2}=0. \label{gem}%
\end{equation}
Then multiplying the equation by an imaginary displacement $dr_{h}$ of the
horizon, and introducing some constants, we can rewrite it as%
\begin{equation}
\underset{k_{B}T_{H}}{\underbrace{\frac{\hbar cN_{t}(r_{h})}{2\pi}\left[
N_{t}^{^{\prime}}(r_{h})-N_{r}^{^{\prime}}(r_{h})\right]  }}\underset
{dS}{\underbrace{\frac{c^{3}}{G\hbar}d\left(  \frac{1}{4}4\pi r_{h}%
^{2}\right)  }}-\underset{dE}{\underbrace{\frac{1}{2}\frac{c^{4}}{G}dr_{h}}%
}=0, \label{gt}%
\end{equation}
and read off the expressions:%
\begin{align}
S  &  =\frac{1}{4l_{P}^{2}}\left(  4\pi r_{h}^{2}\right)  =\frac{A_{H}}%
{4l_{P}^{2}};\nonumber\\
E  &  =\frac{c^{4}}{2G}r_{h}=\frac{c^{4}}{G}\left(  \frac{A_{H}}{16\pi
}\right)  ^{1/2}, \label{gbht}%
\end{align}
where $A_{H}$ is the horizon area and $l_{P}^{2}=G\hbar/c^{3}$. If the source
of Einstein's equation is considered, a term $PdV$ will be added to (\ref{gt}) 
as in Ref. \cite{tp02,ksp07}, which will give an exact and strict form
for the first law of thermodynamics by including the matter's influence. Here
we ignore this term only for the brevity. In particular, the method is
classical although we insert the Planck constant in the expressions of the
temperature and the entropy by hand, which showed that the single quantity
should stem from the quantum statistical mechanics. Along this line, many
different situations has been discussed, which indicates the universality of
this kind of identification, see the review \cite{tp10}.

\subsection{Two-dimensional kinematical connection}

Before applying the thermodynamic identification to acoustic black holes, we
have to be sure whether there is the dynamic connection between the fluid and
gravitational models only by the kinematical analogue. Now we give some direct
kinematical relations between 2D black holes and acoustic black holes, which
were once given in Ref. \cite{mc05} but the dynamics is fixed in advance.

Generally, 4D Einstein gravity can decay into 2D by the method of spherical
reduction if the line element can be written as $ds^{2}=g_{\alpha\beta
}dx^{\alpha}dx^{\beta}+e^{-2\Phi\left(  x^{\alpha}\right)  }d\Omega^{2}$ where
$\alpha,\beta=0,1$ and $\Phi\left(  x^{\alpha}\right)  $ is the dilaton field.
In particular, 2D gravity is also called as dilaton gravity \cite{gkv02}.
Start with the action considered in Ref. \cite{mc05,gkv02},%
\begin{equation}
S_{g}=\frac{1}{2}\int d^{2}x\sqrt{-g}\left(  \Phi R+\lambda^{2}V\left(
\Phi\right)  \right)  , \label{dga}%
\end{equation}
where $\Phi$ is a scalar (the dilaton), $\lambda$ is a parameter for the
balance of the dimension, and $V\left(  \Phi\right)  $ is the dilaton
potential. The equation of motion related to the variable $g_{\alpha\beta}$ is%

\begin{equation}
R=-\lambda^{2}\frac{dV}{d\Phi}.
\end{equation}
The model admits black-hole solutions with the form in Schwarzschild gauge as
\cite{lk94}
\begin{equation}
ds^{2}=-\left(  J\left(  \Phi\right)  -\frac{2M}{\lambda}\right)  d\tau
^{2}+\left(  J\left(  \Phi\right)  -\frac{2M}{\lambda}\right)  ^{-1}%
dr^{2},\Phi=\lambda r, \label{dbh}%
\end{equation}
where $M$ is the mass of the black hole and $J=%
{\displaystyle\int}
Vd\Phi$. The location $r_{h}$ of the black-hole horizon is determined by
$J\left(  \lambda r_{h}\right)  =2M/\lambda$. The metric is also gotten by
taking $N_{t}=1,N_{r}=\sqrt{1-J-\frac{2M}{\lambda}}$ from the ADM form
(\ref{me}), but its relation to the 2D section of Schwarzschild black hole is
subtle, which will be involved later.

Comparing the metric (\ref{dbh}) with the acoustic metric (\ref{abhm}) in two
dimension, the kinematical relation can be gotten as%
\begin{equation}
r\sim\int\rho dx,\tau\sim t+\int dx\frac{v}{c_{s}^{2}-v^{2}},J\sim\frac
{2M}{\lambda}+\frac{\rho}{c_{s}}\left(  c_{s}^{2}-v^{2}\right)  , \label{kr}%
\end{equation}
as presented in \cite{mc05}.

When we endow the thermodynamic-like description to the acoustic black hole,
the role played by the relations (\ref{kr}) will be seen. Now we turn to
another kinematical relation. As stated in Ref. \cite{mv97,mv98}, Hawking
radiation is purely kinematic effect, and so is the temperature. Thus, one can
relate the temperature $T_{b}=\frac{\lambda}{4\pi}V\left(  \lambda
r_{h}\right)  $ for the 2D black holes \cite{rbm93,rcm94} with that in Eq.
(\ref{at}) for the acoustic black hole by%
\begin{equation}
\frac{\hbar}{2\pi k_{B}}v^{^{\prime}}\left(  x_{h}\right)  \sim\frac{\lambda
}{4\pi}V\left(  \lambda r_{h}\right)  , \label{tae}%
\end{equation}
which, together with the fluid's equation at the horizon where the two
equations of motion for the fluid have the same form, gives a constraint for
the fluid dynamics at the horizon. But it has to be stressed that this is not
evident for 4D situation, since the conformal factor (that is irrelevant to
the kinematical results, but can influence the dynamics) is not treated
properly in kinematical analogue there. Therefore, one has to be careful about
4D situation, but at least in two dimension, we can proceed to the discussion
about thermodynamics for acoustic black holes.

\subsection{Correspondence between full actions}

As well-known, the acoustic metric is obtained through the linear
perturbation, so in this subsection we will give a brief proof that at the
perturbative level, the two cases are still equivalent. For the acoustic
fluid, the corrected action is given by Eq. (\ref{abhp}), and in two
dimension, the expression is%

\begin{equation}
S_{P}=-\frac{1}{2}\int d^{2}xA\left[  \rho\left(  \partial_{x}\phi\right)
^{2}-\frac{\rho}{c_{s}^{2}}\left(  \partial_{t}\phi+v\cdot\partial_{x}%
\phi\right)  ^{2}\right]  , \label{abhp2}%
\end{equation}
where $A$ is the cross-sectional area of fluid and $\phi$ is the perturbative
field whose equation of motion is given by
\begin{equation}
\nabla^{2}\phi=\frac{1}{\sqrt{-g}}\partial_{\mu}\left(  \sqrt{-g}g^{\mu\nu
}\partial_{\nu}\phi\right)  =0. \label{abhre}%
\end{equation}

For 2D dilaton gravity, the corrected action can be given by the
Polyakov-Liouville action (see Ref. \cite{gkv02} and references therein),%
\begin{equation}
I_{PL}=-\frac{\alpha}{2\pi}\int d^{2}x\sqrt{-g}\left(  \frac{\left(
\nabla\varphi\right)  ^{2}}{2}+\varphi R\right)  ,
\end{equation}
where $\alpha$ is quantum coupling parameter related to the number of
conformal scalar fields, $R$ is 2D scalar curvature, and $\varphi$ field is
related to the backreaction of spacetime caused by Hawking radiation. The
equation of motion for the $\varphi$ field is,
\begin{equation}
\nabla^{2}\varphi=R. \label{dbhre}%
\end{equation}

Within the Schwarzschild gauge,%
\begin{equation}
ds^{2}=-X\left(  r\right)  d\tau^{2}+\frac{1}{X\left(  r\right)  }dr^{2},
\end{equation}
Eq. (\ref{abhre}) becomes%
\begin{equation}
-\frac{1}{X}\partial_{\tau}^{2}\phi+\frac{dX}{dr}\frac{d\phi}{dr}%
+X\partial_{r}^{2}\phi=0.
\end{equation}
and Eq. (\ref{dbhre}) becomes%
\begin{equation}
-\frac{1}{X}\partial_{\tau}^{2}\varphi+\frac{dX}{dr}\frac{d\varphi}%
{dr}+X\partial_{r}^{2}\varphi=R=-\frac{d^{2}X}{dr^{2}},
\end{equation}
Then we can relate the two equations with $\phi=\varphi+\ln X$. Since $X$ is
independent on the time, the two fields $\phi$ and $\varphi$ have the same
behaviors at the thermodynamic or dynamic level. Furthermore, we can see the
equivalence by the respective numbers of degree of freedom, i.e. for the fluid
model, there are three parameters --- the fluid speed $v$, the fluid density
$\rho$, and the perturbative field $\phi$; for 2D dilaton gravity, there are
also three parameters --- the dilaton field $\Phi$, the metric field
$g_{\mu\nu}$ (only one component in the Schwarzschild gauge), and the
Polyakov-Liouville field $\varphi$. But for 4D situation, there is never so
nice correspondence, and even for the 4D Schwarzschild black holes, the
gravitational parameters can not be modeled by the fluid parameters directly,
which requires a conformal parameters to relate them, but the conformal
parameter is dependent on the time, so one must be careful to treat 4D
situation, in particular at the level of thermodynamics.

\subsection{Thermodynamic identification}

As far as we know, the method of thermodynamic identification has not been
applied for 2D gravity. Here we give a brief implementation for this. At the
horizon of black hole (\ref{dbh}), the field equation for 2D dilaton gravity
model (\ref{dga}) becomes%
\begin{equation}
\frac{dJ}{dr}|_{r_{h}}=\lambda V\left(  \lambda r_{h}\right)  .
\end{equation}
When the imaginary displacement $dr_{h}$ of the horizon is included, the
equation is reexpressed with the form,%
\begin{equation}
\underset{k_{B}T_{b}}{\underbrace{\frac{\hbar\lambda}{4\pi}V\left(  \lambda
r_{h}\right)  }}\underset{dS_{b}}{\underbrace{\frac{2\pi\lambda}{\hbar}dr_{h}%
}}-\underset{dE_{b}}{\underbrace{\frac{\lambda}{2}dJ\left(  \lambda
r_{h}\right)  }}=0.
\end{equation}
From this, one can read off
\begin{align}
S_{b}  &  =\frac{2\pi\lambda}{\hbar}r_{h},\nonumber\\
E_{b}  &  =\frac{\lambda}{2}J\left(  \lambda r_{h}\right)  , \label{dbhse}%
\end{align}
up to an integral constant, respectively. In particular, they are consistent
with the expressions of entropy and energy obtained by using other methods \cite{rbm93,rcm94}.

Now we begin to discuss the thermodynamics for an acoustic black hole. Taking
the derivative with regard to $x$ for the steady Bernoulli equation
(\ref{be}), we have the Euler's equation in two dimension,%
\begin{equation}
\rho v\frac{dv}{dx}+c_{s}^{2}\frac{d\rho}{dx}=0, \label{af}%
\end{equation}
which is one of equations of motion in two dimension but it is consistent with
the other one at the horizon. Considering the expression of the temperature
(\ref{at}), we can rewrite the equation at the horizon as%
\begin{equation}
\underset{k_{B}T_{a}}{\underbrace{\frac{\hbar}{2\pi}v^{^{\prime}}\left(
x_{h}\right)  }}\underset{dS_{a}}{\underbrace{\frac{2\pi}{\hbar K}\rho\left(
x_{h}\right)  dx_{h}}}-\underset{dE_{a}}{\underbrace{\frac{c_{s}}{K}%
d\rho\left(  x_{h}\right)  }}=0,
\end{equation}
where $K$ is introduced in order to balance the dimension. Then we read off
\begin{align}
S_{a}  &  =\frac{2\pi}{\hbar K}\int^{x_{h}}\rho\left(  x\right)
dx;\nonumber\\
E_{a}  &  =\frac{c_{s}}{K}\int^{x_{h}}\rho^{^{\prime}}\left(  x\right)  dx,
\label{abhsm}%
\end{align}
where the integral is made near the horizon. Then we attempt to understand
these identifications (\ref{abhsm}). The mass can be interpreted with a
force-density term in the Euler's equation \cite{blv11}, that is $c_{s}%
^{2}\frac{d\rho}{dx}=\frac{\delta E_{a}}{\delta x}=$ $F_{x}\equiv-\rho
\frac{d\mu}{d\rho}\frac{d\rho}{dx}$. Here it is also regarded as the acoustic
analogue of gravitational mass of black holes. The entropy can be interpreted
as from the property of the horizon, but it can also be interpreted with other
ways, i.e. brick wall model \cite{sg11,mr11,mr13,abs14} that assumes the black
hole is under an equilibrium state with the thermal gas surrounding it. Note
that our identification for the entropy of the acoustic black hole is formally
different from that from brick wall model \cite{mr11,mr13}, but one can obtain a
consistent form from the two methods when the system¡¯s evolution
is considered; that is, $\dot{S_{a}}\varpropto
\kappa$ (that is the surface gravity of acoustic black holes via
$T=\frac{\kappa}{2\pi}$) by using the results of Ref. \cite{bff05} for
backreaction expression of $\rho$ or $v$, in line with \cite{sg11,mr11}.

It is evident that the thermodynamics of acoustic black holes cannot be
Schwarzschild-like, since the Schwarzschild analogue usually takes
$\rho\left(  x\right)  \propto x^{-3/2}$ and $v\left(  x\right)  \propto$
$x^{-1/2}$ up to a conformal difference. One might suspect that this
inconsistency is due to the one-dimensional linear fluid model that we take
approximately (see the discussion below the metric (\ref{abhm})), but the same
mathematical forms will also be gotten by reducing the action (\ref{abha})
with the consideration of spherically symmetry, since $\overrightarrow{\nabla
}\psi=\frac{\partial\psi}{\partial x}\hat{x}+\frac{\partial\psi}{\partial
y}\hat{y}+\frac{\partial\psi}{\partial z}\hat{z}=\frac{\partial\psi}{\partial
r}\hat{r}+\frac{1}{r}\frac{\partial\psi}{\partial\theta}\hat{\theta}+\frac
{1}{r\sin\theta}\frac{\partial\psi}{\partial\phi}\hat{\phi}$ that reduces to
$\overrightarrow{\nabla}\psi=\frac{\partial\psi}{\partial x}\hat{x}$ for the
case discussed in the paper or $\overrightarrow{\nabla}\psi=$ $\frac
{\partial\psi}{\partial r}\hat{r}$ when the velocity is suppressed in two
angular directions. Thus, if a two-dimensional fluid model is considered, it
might not related to the simple reduction of four-dimensional gravity,
described by Eq. (\ref{gem}), so its thermodynamics doesn't have to be Schwarzschild-like.

As a consistent check, we will see that the kinematical relation (\ref{kr})
can give the thermodynamic connection between the 2D dilaton black hole and
the acoustic black hole, i.e. taking $\lambda\sim\frac{1}{K}$, the relation
$r\sim\int\rho dx$ indicates $S_{b}\sim S_{a}$; from the relation $J\sim
\frac{2M}{\lambda}+\frac{\rho}{c_{s}}\left(  c_{s}^{2}-v^{2}\right)  $, one
has $dJ\sim$ $2\rho dv\sim$ $2c_{s}d\rho$ at the horizon, which indicates
$E_{b}\sim E_{a}$. Therefore, the kinematical relations leads to a direct
connection between the thermodynamics of 2D dilaton black hole and acoustic
black hole, under the condition that any relation between fluid dynamics and
gravitational dynamics is not assumed in advance.

In order to present the universal property of thermodynamic identification
(\ref{abhsm}), we will also include the external potential explicitly in the
discussion of acoustic black hole with another kind of popular model that is
related to BEC. For the analogue of BEC, the corresponding Euler equation can
be expressed as \cite{gaz00,blv01},%
\begin{equation}
\rho_{B}v_{B}\frac{dv_{B}}{dx}+\frac{\rho_{B}}{m}\frac{dV_{ext}}{dx}+c_{B}%
^{2}\frac{d\rho_{B}}{dx}=0. \label{abheu}%
\end{equation}
This can be obtained from the Gross-Pitaevski equation, $i\hbar\partial
_{t}\Psi=\left(  -\frac{\hbar^{2}}{2m}\nabla^{2}+V_{ext}+\frac{4\pi a\hbar
^{2}}{m}\left\vert \Psi\right\vert ^{2}\right)  \Psi$ where the condensate is
considered in the dilute gas approximation and nearly all atoms are in the
same single-particle quantum state $\Psi\left(  x,t\right)  $, $m$ is the mass
of individual atoms, $a$ is the scattering length, and $V_{ext}\left(
x\right)  $ is the external potential that trapped these bosons. In
particular, if a background stationary state, $\Psi_{B}\left(  x,t\right)
=\sqrt{\rho_{B}\left(  x\right)  }e^{i\psi_{B}\left(  x\right)  }e^{-i\omega
t}$, is considered, the propagation of small perturbations of the condensate
around this background can be calculated to get the acoustic metric, as
presented in Eq. (\ref{abhm}). The velocity of the background is given by
$v_{B}=\frac{\hbar}{m}\nabla\psi_{B}\left(  x\right)  $, and the local
velocity of sound by $c_{B}\left(  x\right)  =\frac{\hbar}{m}\sqrt{4\pi
a\rho_{B}\left(  x\right)  }$. Moreover, the quantum pressure term $Q\left(
\overrightarrow{x}\right)  =-\frac{\hbar^{2}}{2m}\frac{\nabla^{2}\sqrt
{\rho_{B}\left(  \overrightarrow{x}\right)  }}{\sqrt{\rho_{B}\left(
\overrightarrow{x}\right)  }}$ is ignored since BEC for the analogue of
acoustic black holes always works within the regime of validity of
Thomas-Fermi approximation \cite{dgs99} that the condensate does not vary on
length scales shorter than the healing length $\xi=1/\sqrt{8\pi\rho_{B}a}$.

With the temperature of the acoustic black hole known in advance (note that
the speed of sound here is not constant), the Eq. (\ref{abheu}) can be
rewritten at the horizon as%
\begin{equation}
\underset{k_{B}T}{\underbrace{\frac{\hbar}{2\pi}\left(  v_{B}^{^{\prime}%
}\left(  x_{h}\right)  +c_{B}^{^{\prime}}\left(  x_{h}\right)  \right)  }%
}\underset{dS_{B}}{\underbrace{\frac{2\pi}{\hbar K}\rho_{B}\left(
x_{h}\right)  dx}}-\underset{dE_{B}}{\underbrace{\left(  \frac{\rho_{B}\left(
x_{h}\right)  }{Kmc_{B}\left(  x_{h}\right)  }dV_{ext}\left(  x_{h}\right)
+\frac{3c_{B}\left(  x_{h}\right)  }{2K}d\rho_{B}\left(  x_{h}\right)
\right)  }}=0.
\end{equation}
It is easily seen that the identification of the entropy is the same with that
in Eq. (\ref{abhsm}). The mass is expressed as,
\begin{equation}
E_{B}=\frac{1}{K}\int^{x_{h}}\left[  \frac{\rho_{B}\left(  x_{h}\right)
}{mc_{B}\left(  x_{h}\right)  }dV_{ext}\left(  x_{h}\right)  +\frac
{3c_{B}\left(  x_{h}\right)  }{2}d\rho_{B}\left(  x_{h}\right)  \right]
\end{equation}
where the expression of $E_{B}$ is different from that of $E_{a}$ due to the
non-constant speed of sound and the external potential. But it is noted that
the expression of entropy is model-independent. This can also be understood by
the entropic gravity \cite{ev11}, which assumed that the gravity is derived
from the change of entropy with the form $\Delta S=2\pi k_{B}\frac{\Delta
x}{\lambdabar}$. Thus if we take the proper form for the constant $K$ and the
relation $r\sim\int\rho dx$, the entropy in Eq. (\ref{abhsm}) is consistent
with $\Delta S$, which show again that the thermodynamic-like description for
analogous black hole is feasible at least in 2D.

\section{Specific heat}

From the discussion above, we have known the expressions of thermodynamic
quantities of acoustic black holes which conform to the first law of
thermodynamics. Actually, more importantly, we want to discuss how to use
these quantities to describe the evolution of acoustic physical systems, i.e.
the change caused by emission of thermal radiation from the acoustic horizon.
The calculation of backreaction provided a fundamental method to answer this
question, but here we want to estimate it via specific heat of an acoustic
black hole, which will give the information about the change of temperature
during the radiation. Generally, the temperature in black hole theory is a
geometric quantity related closely to the spacetime background, so the change
of temperature will indicate the change of spacetime background that is also
called as backreaction if the change is not so violent.

According to our results in Eqs. (\ref{at}) and (\ref{abhsm}), the specific
heat of an acoustic black hole can be written as%
\begin{equation}
C_{a}=\frac{dE_{a}}{dT_{a}}=\frac{dE_{a}/dx}{dT_{a}/dx}=-\frac{2\pi k_{B}%
\rho\left(  x_{h}\right)  }{\hbar K}\frac{v^{^{\prime}}\left(  x_{h}\right)
}{v^{^{\prime\prime}}\left(  x_{h}\right)  }. \label{tsh}%
\end{equation}
It is easily seen that the sign of specific heat is dependent on the first and
second derivatives of the velocity with regard to the coordinate $x$. If we
take the model of Laval nozzle that is described in Ref. \cite{bff05}, which
gave $v^{^{\prime}}\left(  x_{h}\right)  >0$, and $v^{^{\prime\prime}}\left(
x_{h}\right)  <0$, we find a positive specific heat. This means that the
temperature will decrease after the radiation is emitted, and such behavior of
acoustic black holes resembles a near-extremal Reissner-Nordstr\"{o}m black
hole while not a Schwarzschild black hole, which is consistent with the result
of backreaction analysis \cite{bff05}. Thus, it indicates that the
thermodynamics of acoustic black holes is model-dependent. The extension to the case where the speed
of sound is not constant is possible and will not change our conclusion about
the model-dependent thermodynamic-like behaviors for the corresponding
acoustic black holes.

The specific heat of 2D dilaton black holes is also easy to be gotten,%
\begin{equation}
C_{b}=\frac{dE_{b}}{dT_{b}}=2\pi\frac{V\left(  \Phi_{h}\right)  }{dV\left(
\Phi_{h}\right)  /d\Phi}, \label{dsh}%
\end{equation}
which is consistent with that obtained in Ref. \cite{mc97,mkp07}. In
particular, the specific heat is dependent on the dilaton potential which can
lead to a model similar to the near-extremal Reissner-Nordstr\"{o}m black hole
under some conditions. Then, whether the two specific heats have any relation
under the kinematical analogue. A straight use of kinematical relations in Eq.
(\ref{tae}) and $r\sim\int\rho dx$ shows $C_{a}\sim C_{b}$, which is a further
evidence for the connection between the two dynamics with only the kinematical
analogue made in advance.

From the analysis above, it is seen that the derivative of the parameter $v$
in the fluid can be regarded as the analogue of dilaton potential, which is
equivalent to such analogue, $v\sim M$, where $M$ is the mass of 2D dilaton
black hole. This is different from the usual kinematical analogue for 2D
sections of Schwarzschild black hole, that is $v\sim\sqrt{\frac{2M_{H}}{r_{H}%
}}$, where $M_{H},r_{H}$ are the mass and the radial coordinate of
Schwarzschild black hole, respectively. In fact a specific situation for 2D
dilaton black hole, i.e. $V\left(  \Phi\right)  =1/\sqrt{2\Phi},$ can be taken
in the Schwarzschild form by making the transformation $g_{\mu\nu}%
\rightarrow\frac{1}{\sqrt{2\Phi}}g_{\mu\nu},r\rightarrow\frac{\lambda}{2}%
r_{H}^{2}$ for the metric (\ref{dbh}). This means the kinematical relation in
Eq. (\ref{tae}) becomes $\frac{\hbar}{2\pi k_{B}}v^{^{\prime}}\left(
x_{h}\right)  \sim\frac{\lambda}{4\pi}\frac{1}{\sqrt{2\Phi_{h}}}\sim
\frac{\hbar}{8\pi Gk_{B}M}$ where the speed of light is taken as $c=1$. As
discussed in the last section, the kinematical relations can lead to the
relation between the mass of acoustic black hole and gravitational black hole,
so we have $v^{^{\prime}}\left(  x_{h}\right)  \sim\frac{1}{4GE_{a}}\sim
\frac{1}{-4\int^{x_{h}}\rho\left(  x\right)  v^{^{\prime}}\left(  x\right)
dx}$. Thus the relation for the parameter $v$ in the fluid is%
\begin{equation}
-4v^{^{\prime}}\left(  x_{h}\right)  \int^{x_{h}}\rho\left(  x\right)
v^{^{\prime}}\left(  x\right)  dx\sim1, \label{rtbh}%
\end{equation}
which is necessary to model the thermal behavior of a Schwarzschild black
hole. With this relation, it is easy to confirm that $C_{a}<0$. Of course, one
can also obtain a similar relation for the other physical systems like BEC to
attempt to find the Hawking-like radiation as emitted from a Schwarzschild
black hole. Further, as seen from the transformation for 2D Schwarzschild
metric into dilaton black-hole metric, a conformal factor related to the
dilaton potential is involved. Thus it is understood that why the acoustic
black holes always presents directly the behaviors of 2D dilaton black holes,
while an extra constraint is required for the simulation of Schwarzschild-like
black holes.

\section{Conclusion}

In this paper, we have reinvestigated the concept of acoustic black holes and
discussed background fluid equations. Even though the fluid equations cannot
give the complete expressions for each field without any extra knowledge
besides the equations of motion, they still include a wealth of information.
Via fluid's equation of motion as a thermodynamic identity, we have read off
the mass and the entropy for acoustic black holes with the temperature known
in advance. This identification is guaranteed by the kinematical connection
which can lead to the dynamic connection between the fluid and gravitational
models in two dimension. In particular, through the analysis for the
kinematical relations, we have found that the thermodynamics of acoustic black
holes reproduces the thermodynamics of two-dimensional dilaton black holes
exactly, and so the fluid can be regarded as a natural analogue of
two-dimensional dilaton gravity, which is significant for many ongoing related
experimental observations. Novelly, we have also found that the entropy for
acoustic black holes is model-independent and have an interpretation similar
to the entropic gravity. Moreover, it is found that the derivative of velocity
$v$ of the background fluid is a nice analogue of the dilaton potential, from
the kinematical and thermodynamic perspective, respectively, which also
interprets why the 4D kinematical analogue is difficult to lead to any dynamic
analogue, due to conformal difference.

With the mass and the temperature identified by thermodynamics, we have
proceeded to get the specific heat for the acoustic black hole, and found the
sign of the specific heat is model-dependent, which means that some extra
knowledge besides the equations of motion must be given to estimate the
thermodynamic stability. However, when we want to model some kind of black
holes, i.e. Schwarzschild black holes, the relation about the parameters in
the fluid can be known only by the fluid equation and the thermodynamic
correspondence, as made in Eq. (\ref{rtbh}). Finally, all our analysis is made
for 2D model, so whether they can be extended to 4D or higher dimensions is
unclear now. It might be possible, however, to relate 2D with 4D cases by taking into account properly
the behaviour of 2D dilaton gravity under Weyl rescaling of the metric \cite{mc97}, which deserves a further investigation.

\section{Acknowledgements}

The author acknowledges the support from Grant No. 11374330 of the National
Natural Science Foundation of China and from Open Research Fund Program of the
State Key Laboratory of Low-Dimensional Quantum Physics and from the
Fundamental Research Funds for the Central Universities, China University of
Geosciences (Wuhan) (No. CUG150630).

\end{document}